# *FDTD Analysis of the Tunneling and "Growing Exponential" in a Pair of ε-negative and μ-negative Slabs*

Andrea Alù[1,2], Nader Engheta[1] and Richard W. Ziolkowski[3]

*[1]University of Pennsylvania*
*Department of Electrical and Systems Engineering*
*Philadelphia, Pennsylvania 19104, U.S.A.*
*engheta@ee.upenn.edu, http://www.ee.upenn.edu/~engheta*

*[2]University of Roma Tre*
*Department of Applied Electronics, Rome, Italy*
*alu@uniroma3.it, http://www.dea.uniroma3.it/lema/people/andrea_alù.htm*

*[3]University of Arizona*
*Electrical and Computer Engineering Department, Tucson, Arizona, U.S.A.*
*ziolkowski@ece.arizona.edu*

## ABSTRACT

Pairing together material slabs with opposite signs for the real parts of their constitutive parameters has been shown to lead to interesting and unconventional properties that are not otherwise observable for single slabs. One such case was demonstrated analytically for the "conjugate" (i.e., complementary) pairing of infinite planar slabs of ε-negative (ENG) and μ-negative (MNG) media [A. Alù, and N. Engheta, *IEEE Trans. Antennas Prop.*, 51, 2558 (2003)]. There it was shown that when these two slabs are juxtaposed and excited by an incident plane wave, resonance, complete tunneling, total transparency and reconstruction of evanescent waves may occur in the steady-state regime under a monochromatic excitation, even though each of the two slabs by itself is essentially opaque to the incoming radiation. This may lead to virtual imagers with sub-wavelength resolution and other anomalous phenomena overcoming the physical limit of diffraction. Here we explore how a transient sinusoidal signal that starts at $t = 0$ interacts with such an ENG-MNG pair of finite size using an FDTD technique. Multiple reflections and transmissions at each interface are shown to build up to the eventual steady state response of the pair, and during this process one can observe how the "growing exponential" phenomenon may actually occur inside this bilayer.

## INTRODUCTION

The current interest in understanding the physics behind the anomalous properties of metamaterials is evident in the recent physics and engineering literature. In particular, artificial materials with negative constitutive parameters, which can be distinguished into ε-negative (ENG), μ-negative (MNG) [1] and double-negative (DNG) [2] media if, respectively, their effective permittivity, permeability, or both of them have a negative real part, have been at the center of this attention, due to the anomalous phenomena theoretically predicted for their behavior and, in part, also verified experimentally. The existence and the possible artificial realization of these materials have been studied and verified in the past and recent times. In particular, ENG (plasmonic) media exist naturally in the infrared and optical frequencies, e.g., noble metals below their plasma frequency [3] and polar dielectrics, and they can be relatively easily synthesized at lower frequencies by embedding a regular lattice of thin metallic wires in a host medium [4]. These inclusions provide the proper resonant electric polarizability in such an artificial material in the desired frequency regime. In analogy, an MNG material, such as a resonant ferro-magnetic medium, may be synthesized by embedding resonant magnetic loops in a host medium, i.e., split-ring resonators, thereby providing the proper magnetic resonant polarizability in the desired frequency regime [5]. These two techniques may be employed at the same frequency to obtain DNG materials within a given frequency range, as has been reported in [6]-[7]. To be consistent with the previous terminology, we will refer in the following to common materials, which have both positive permittivity and permeability, as double-positive (DPS) materials.

One of the most striking properties of a planar DNG slab is represented by the possibility that it can focus sub-wavelength details, as first predicted in [8], by fostering the "growth" (instead of the decay) of evanescent waves inside it. Following this discovery, a multitude of papers explaining the potential implications and possible limitations of this anomalous phenomenon when realistic metamaterials are considered have appeared (see e.g., [9]-[14]). In relation to this phenomenon it was demonstrated analytically that the "conjugate" (i.e., complementary) pairing of infinitely extent in two dimensions, juxtaposed planar slabs of ENG and MNG media [1] (as well as DPS and DNG pairs, of which the *perfect* lens [8] is a special case), may induce an anomalous resonance, complete tunneling, total transparency and reconstruction of evanescent waves, even though each of the two slabs by itself is essentially opaque to the incoming radiation.

Most of these studies, however, have been performed analytically or numerically for the time-harmonic steady-state regime, i.e., under monochromatic excitation at a given fixed frequency, at which the involved metamaterials were supposed to have an ENG, MNG or DNG response. However, the dispersive nature of these materials, which is well known to be necessary for energy conservation issues in passive materials with negative constitutive parameters [15], requires a certain time to establish this desired steady-state response as demonstrated previously [16]. Moreover, due to the resonant nature of the phenomena associated with these juxtaposed slabs and its connection to surface-plasmons, it is clear that a certain amount of time is required to establish this behavior and it would be inversely related to the Q factor of this resonance. It is therefore expected that the response of these slab structures to a realistic initial excitation is not instantaneous, but would rely on some number of multiple reflections at each interface to establish the resonant mode.

In the following, we explore how a transient sinusoidal signal interacts with a juxtaposed ENG-MNG pair, showing how the multiple reflections and transmissions at each interface can be designed to build up to the eventual resonant steady-state response of the pair. We will demonstrate that one can observe the "growing exponential" phenomenon in this process as it occurs inside this bilayer. Here we have simulated this time-domain problem using the finite-difference time-domain (FDTD) technique, assuming the Drude models for the frequency dependence of the permittivity and permeability of these slabs. This choice allowed us to incorporate dispersive effects into the simulation model along with the possibility of having the material properties of both of these regions attain specified negative real parts at or approximately near the frequency of the sinusoidal excitation. Our time-domain results confirm the steady-state prediction of "growing exponential" behavior in the bilayer and the total resonant transmission through it. However as anticipated, these responses are achieved only after a certain period of time (i.e., a certain necessary number of FDTD time steps), which allows for the buildup of the interactions between the interfaces. Consequently, this time delay is found to be in general a function of the thickness of the slabs and the possible losses in each slab. A sketch of a small portion of these results was presented in a recent symposium [17].

## FORMULATION OF THE PROBLEM

Two classes of numerical simulations were considered here: one-dimensional (1D) and two-dimensional (2D). The 1D problem represents a plane-wave normally incident on a pair of infinitely extent ENG-MNG slabs. The two slabs have, respectively, the permittivity $\varepsilon_{ENG}$, $\varepsilon_{MNG}$, the permeability $\mu_{ENG}$, $\mu_{MNG}$, and the thicknesses $d_{ENG}$, $d_{MNG}$ in the direction of propagation.

The geometry of the 2D numerical problem is depicted in Fig. 1. Two planar slabs of transverse length $l$ are surrounded by the empty space (with permittivity $\varepsilon_0$ and permeability $\mu_0$) in a suitable Cartesian reference system. These two slabs also have, respectively, the permittivities $\varepsilon_{ENG}$, $\varepsilon_{MNG}$, the permeabilities $\mu_{ENG}$, $\mu_{MNG}$, and the thicknesses $d_{ENG}$, $d_{MNG}$. The excitation in the figure represents a TM (parallel polarized) plane wave impinging at a specified angle $\theta_i$ with respect to the normal. For simplicity, the geometry is independent of $y$. The actual excitation in the 2D simulations was a TM Gaussian beam; this provided us with the means to study the localization of the fields in the bilayer and beyond it.

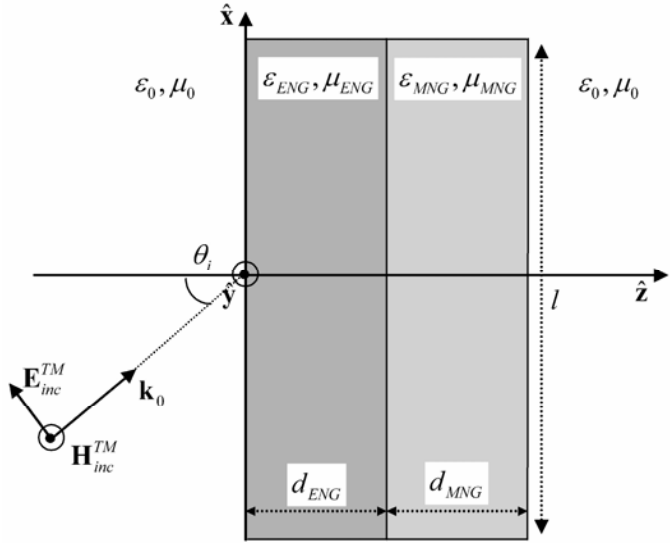

Fig. 1. A juxtaposed pair of planar finite sized slabs that are located in free space and are excited by a TM plane wave.

When the materials have negative parameters, the constitutive parameters of the two materials composing the slabs under both the 1D and 2D analyses are assumed to follow the lossy Drude models for their frequency dependence [2]. Their general form for a $e^{-i\omega t}$ time dependence may be written as:

$$\varepsilon_{ENG}(\omega) = \varepsilon_0 \left[1 - \frac{\omega_{peENG}^2}{\omega(\omega + i\Gamma_{eENG})}\right], \mu_{ENG}(\omega) = \mu_0 \left[1 - \frac{\omega_{pmENG}^2}{\omega(\omega + i\Gamma_{mENG})}\right]$$
$$\varepsilon_{MNG}(\omega) = \varepsilon_0 \left[1 - \frac{\omega_{peMNG}^2}{\omega(\omega + i\Gamma_{eMNG})}\right], \mu_{MNG}(\omega) = \mu_0 \left[1 - \frac{\omega_{pmMNG}^2}{\omega(\omega + i\Gamma_{mMNG})}\right]. \quad (1)$$

The electric and magnetic plasma frequencies $\omega_p$ may be chosen so that the two materials are, respectively, an ENG and an MNG medium at the driving radian frequency $\omega_d$, at which the sinusoidal signal is launched. The Drude models represented by (1) satisfy the Kramers-Kronig and causality conditions, as widely discussed in the literature (see e.g., [15]), and also the energy conservation requirements for passive ENG and MNG materials. Notice that these models include the presence of losses, represented by the collision frequency parameters $\Gamma$, and have a limited band of frequencies in which their constitutive parameters are negative, consistent with the limitations exhibited by real-life metamaterials. These models therefore represent valid choices for the following analyses.[1] Note that when the materials are positive and not less than one, their permittivity and permeability parameters are simply taken to be positive constants.

As was shown analytically in [1] for infinite transverse width slabs, a $\sin(\omega_d t)$-monochromatic excitation of such a ENG-MNG pair structure would highlight its anomalous response at the frequency $\omega_d$ if the two media have *complementary* properties. Because the slabs under consideration here are finite in width, we expect similar behavior to occur from this bilayer when the same choice of the geometric and constitutive parameters is made provided that the transverse dimension of the pair is large enough, i.e., $l \gg d_{ENG}, d_{MNG}$. With satisfaction of these conditions, we expect to achieve total tunneling of the radiation through the bilayer with little or no reflection, as well as the "growing" evanescent fields inside this pair and the subsequent possibility of virtual imaging with sub-wavelength resolution.

It is interesting to note that each of the two slabs *per se* would not allow propagation at the excitation frequency, since the wave number $k = \omega_d \sqrt{\mu\varepsilon}$ would be imaginary in either of them. Therefore, it is expected that the system would be highly reflective if one of the two slabs was removed. On the other hand, for the given TM plane wave impinging on the pair at the angle $\theta_i$, the following conditions, which were derived in ([1], eq. 9), lead to the interface resonance:

---

[1] We note that for the negative permeability, Lorentzian or two-time-derivative Lorentzian dispersion models have also been used in the literature [6,7]. Here, however, for the sake of mathematical simplicity and faster convergence in the numerical simulations, the Drude model is used for both negative permittivity and permeability. This choice does not affect the general conclusions and nature of the results reported here.

$$\theta_i = \arcsin\sqrt{\frac{\varepsilon_{ENG}\varepsilon_{MNG}\left(\varepsilon_{MNG}\mu_{ENG} - \varepsilon_{ENG}\mu_{MNG}\right)}{\mu_0\varepsilon_0\left(\varepsilon_{MNG}^2 - \varepsilon_{ENG}^2\right)}}\,,\ \sqrt{\varepsilon_{ENG}\mu_{ENG}}\,d_{ENG} = \sqrt{\varepsilon_{MNG}\mu_{MNG}}\,d_{MNG}\,, \quad (2)$$

These conditions ensure total tunneling, zero reflection and complete phase and amplitude restoration between the entrance and exit face of the bilayer, in the limit of infinitely wide slabs ($l \to \infty$) and no losses. Similar expressions may be derived for the other polarization by invoking duality. An ENG-MNG pair illuminated by an incident plane wave which is designed following the resonance conditions in Eq. (2) will be denoted in the following discussion as a *conjugate pair* (or complementary pair), in analogy with [1].

As a special case, a *conjugate-matched pair* will denote the bilayers in which:

$$\varepsilon_{ENG} = -\varepsilon_{MNG}\,,\ \mu_{ENG} = -\mu_{MNG} \text{ and } d_{ENG} = d_{MNG}\,. \quad (3)$$

These conditions were shown in [1] to guarantee zero-reflection and total-transmission conditions for any TE or TM plane wave impinging on the pair, i.e., independent of the polarization and the angle of incidence. Moreover, when evanescent waves impinge on such conjugate-matched pair, their incident amplitudes are restored as well on the back side of the bilayer. In [1] it was therefore further speculated that such conjugate-matched pairs may act as a virtual image displacer with sub-wavelength resolution, analogous in several ways to the perfect lens presented in [8].

## TRANSIENT RESPONSE: FDTD SIMULATIONS

In this section we verify the analytical predictions reported in [1] and briefly discussed in the previous section with selected numerical simulations generated with an FDTD engine (as in [2]). A sinusoidal excitation at the frequency $\omega_d$ that is smoothly turned on at $t = 0$ is considered for both the 1D and 2D simulations. The spectrum of such an excitation is composed of an infinite number of harmonics centered on the frequency $\omega_d$; this allows one to study the initial transient response of the system and its convergence towards the steady-state conditions.

As a first example, we used the 1D simulator to study the behavior of a bilayer that was designed to be conjugate-matched pair at the frequency $\omega_d$. The material parameters for this case were: $\varepsilon_{ENG}(\omega_d) = -\varepsilon_{MNG}(\omega_d) = -3\varepsilon_0$, $\mu_{ENG}(\omega_d) = -\mu_{MNG}(\omega_d) = 2\mu_0$, $d_{ENG} = d_{MNG} = \lambda_0/10$, with $\lambda_0 = 2\pi/(\omega_d\sqrt{\varepsilon_0\mu_0}) = 1.0 cm$ being the wavelength at the sinusoidal excitation frequency $f_0 = 30 GHz$. The problem space was 10,000 cells long, where $\Delta z = 10\,\mu m = \lambda_0/1000$. The source plane, a total field / scattered field boundary, was located at $z = 2000\Delta z$, the bilayer began at $z = 5000\Delta z$, $3\lambda_0$ away from the source. The thicknesses of the ENG and MNG layers were $d_1 = d_2 = 100\Delta z = \lambda_0/10$. The time step was $\Delta t = 0.95\Delta z/c = 31.67\,fs$.

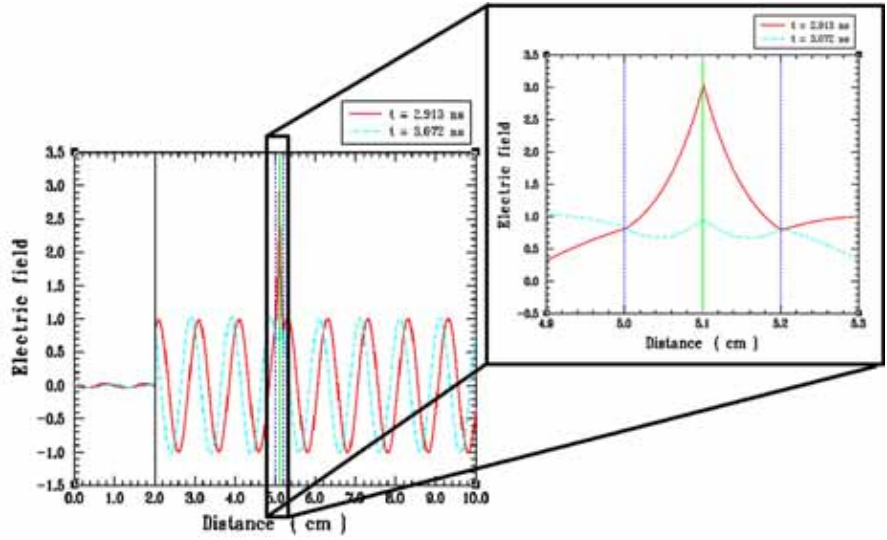

Fig. 2 – 1D FDTD simulation of a conjugate-matched pair
($\varepsilon_{ENG}(\omega_d) = -\varepsilon_{MNG}(\omega_d) = -3\varepsilon_0$, $\mu_{ENG}(\omega_d) = -\mu_{MNG}(\omega_d) = 2\mu_0$, $d_{ENG} = d_{MNG} = \lambda_0/10$)

Fig. 2 shows the electric field distribution inside and outside the bilayer, with a zoom for the distribution inside the conjugate matched pair in the figure inset, at two different, but close snapshots in time when steady-state has already been achieved. The plots clearly show the total tunneling predicted in [1], with the same phase at the entrance and the exit face of the bilayer. Moreover, it is evident how at this point in time the "growing-exponential" distribution is already present, consistent with the fact that the wave is evanescent in each of the slabs of the bilayer, but its amplitude and phase is the same at the entrance and exit face. Notice the sinusoidal variation in time of the exponential peak at the interface between the two layers, which is more evident in the zoom. As mentioned above, here and in the following plots some cells in the entrance side of the bilayer (here the first 2000 cells) are devoted to show only the reflected (scattered) field, isolated from the impinging excitation. As is evident, its amplitude is extremely low in this case, showing the total transmission typical of the steady-state regime for this conjugate-matched pair configuration.

Increasing the thicknesses of the layers to $d_{ENG} = d_{MNG} = \lambda_0/5$, one finds that it takes a longer period of time to reach the steady-state regime. A snapshot of the steady state electric field at $t = 9.476\,ns$, a time comparable in the steady state regime to those given in Fig. 2, is shown in Fig. 3. Again, the time elapsed from $t = 0$ is long enough for the field inside the bilayer to experience the growing exponential, even though the level of reflection here is higher and the transmission amplitude is not complete.

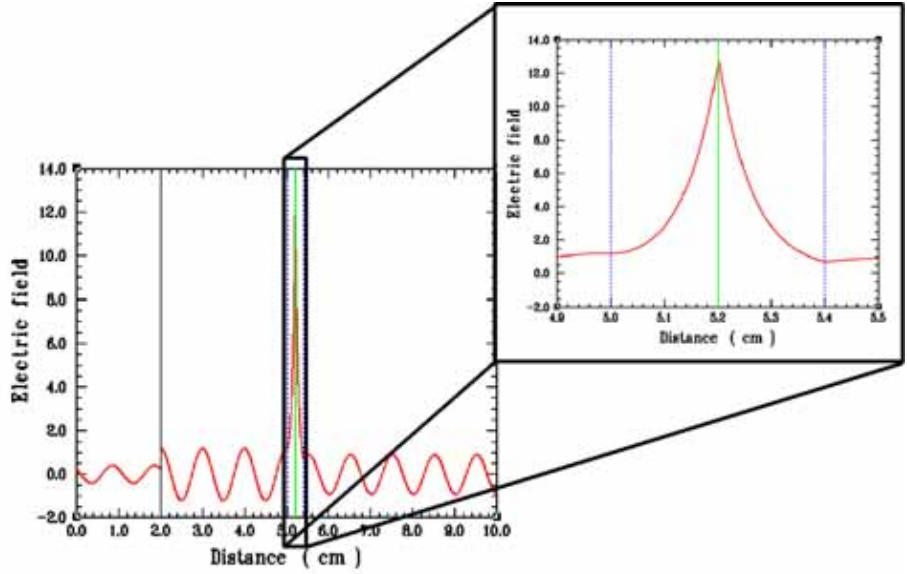

Fig. 3 – FDTD simulation of a conjugate-matched pair (the same as in Fig. 2, but with $d_{ENG} = d_{MNG} = \lambda_0/5$)

Fig. 4 shows the time history for the electric field values at the entrance (red line) and exit (blue line) faces and at the interface (green line) between the two media for the cases of Fig. 2 and 3. Here it is evident how the field at the entrance face rapidly converges to unity, whereas the interface field, which converges to a higher value due to the growth of the exponential predicted theoretically, requires a longer time to reach the steady-state, consistent with the prediction that the multiple reflections inside the bilayer needs some time to build up and achieve a final "growing-exponential" distribution inside the bilayer. This behavior is particularly apparent in the zoom of the first nanosecond in the two figures. It is also consistent with the fact that the spectrum of the excitation is converging towards $\omega_d$, but in the transient the bilayer is not acting as a conjugate-matched pair, since its constitutive parameters have different responses at the different frequency components of the excitation. With a similar behavior, also the exit field converges slowly to unity, showing the total tunneling only after a transient period. Comparing the two cases of Fig. 4, it is clear how the steady-state regime is reached more slowly when thicker slabs are considered, since a larger thickness essentially corresponds to a higher resonance Q factor.

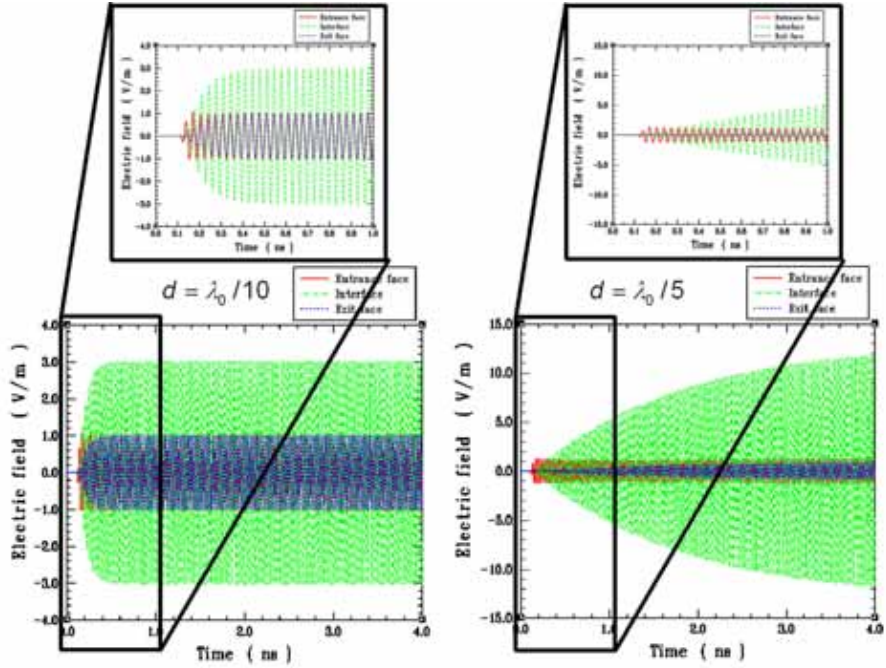

Fig. 4 – Time histories of the field values for the cases of Fig.2 and 3 at the entrance and exit faces and at the interface.

These conclusions were further verified by considering the 2D case of a very broad Gaussian beam that was normally incident on a bilayer, each slab having the same depth as in the 1D examples. The distribution of the Gaussian beam in the total field-scattered field plane was $\exp(-x^2/w_0^2)$ where $w_0 = 5\lambda_0$. The time domain results were essentially the same. There was no noticeable impact on the rate of growth of the interface field of the finite transverse dimension of the slabs, and therefore these results are not reported here.

In Fig. 5 we show the results of a Gaussian beam with $w_0 = 2\lambda_0$ that is obliquely incident on the conjugate-matched bilayer having $d_1 = d_2 = \lambda_0/10$. The angle of incidence of the Gaussian beam was $\theta_i = \pi/9 = 20°$; the total transverse size of the slabs is $8\lambda_0$. The distribution of the electric field intensity is shown for three different instants in time. In the first snapshot, the beam has just arrived on the slab, and it starts its interaction with the bilayer. In the second snapshot, its interaction has started, but we are not yet in the steady-state regime. Here the fields start to build up at the interface between the slabs, while a visible, non-negligible reflected wave occurs and consequently creates an interference pattern with the incident wave. Entering the steady-state regime (third snapshot), however, the bilayer becomes totally transparent to the radiation, the fields are sensibly higher at the interface than in the outside region, and the reflection from the bilayer is zero. Notice also how the phase of the plane wave is totally restored at the exit face of the bilayer, as if the structure were completely transparent to the plane wave incidence.

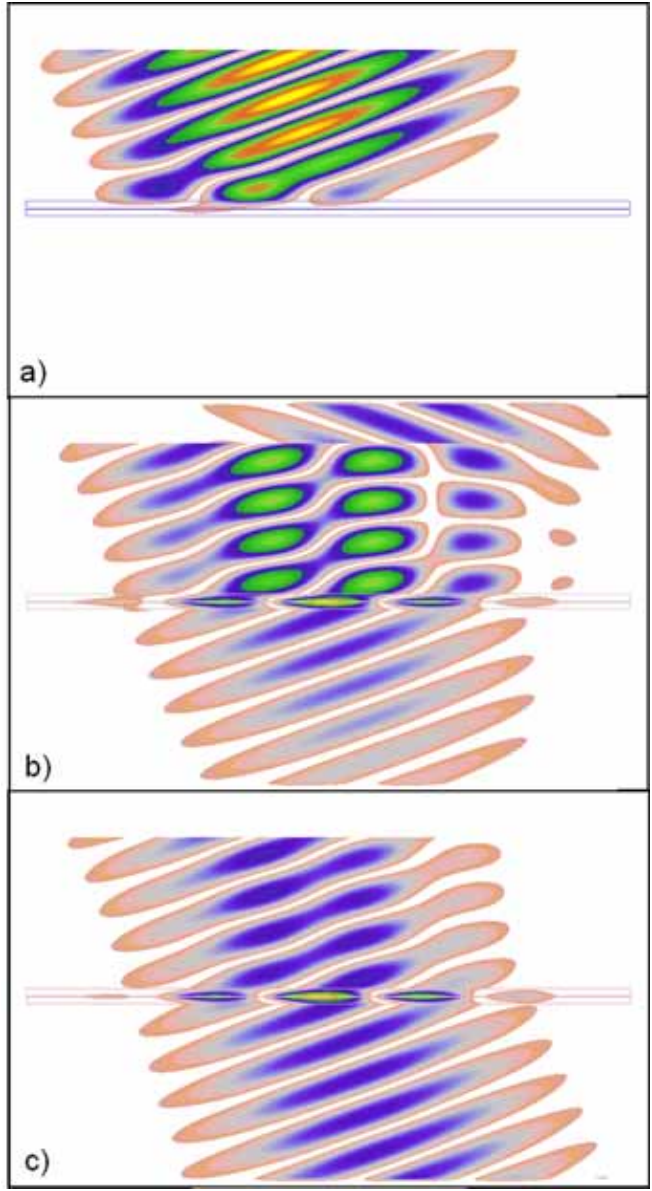

Fig. 5 – Three different snapshots in time (a) $t = 600\Delta t = 134.3\,ps$, (b) $t = 1200\Delta t = 268.7\,ps$, and (c) $t = 3800\Delta t = 850.8\,ps$, for a Gaussian beam with $w_0 = 2\lambda_0$ incident at the angle $\theta_i = \pi/9 = 20°$ on the conjugate-matched pair of Fig. 2.

The same behavior was also verified for a Gaussian beam with $w_0 = 0.5\lambda_0$ impinging with $\theta_i = \pi/9 = 20°$ on the same conjugate-matched bilayer. These results are shown in Fig. 6. Since we are considering here, as in Fig. 5, a conjugate-matched bilayer, which in the steady-state is transparent at every angle of incidence, we verify again in the last snapshot given in Fig. 6 that total tunneling occurs through the bilayer even for this more complex excitation. Again, note that the phase-restoration phenomenon is evident in the structure. In fact, one can see that the Gaussian beam actually tunnels in phase and amplitude through the bilayer when the steady state is reached.

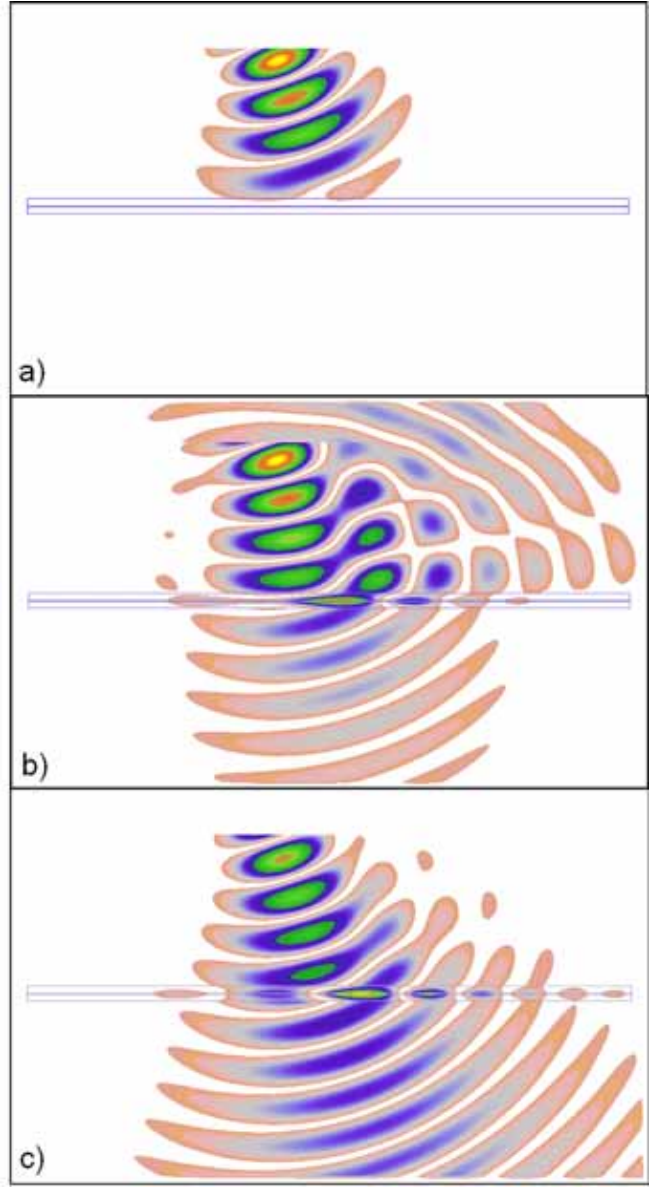

Fig. 6 – Three different snapshots in time (a) $t = 600\Delta t = 134.3\,ps$, (b) $t = 1200\Delta t = 268.7\,ps$, and (c) $t = 3600\Delta t = 806.0\,ps$, for a Gaussian beam with $w_0 = 0.5\lambda_0$ incident at the angle $\theta_i = \pi/9 = 20°$ on the conjugate-matched pair of Fig. 2.

We have also considered just a conjugate bilayer, which is designed following Eq. (2) to achieve the anomalous total tunneling only at a specific incidence angle. In this case, we have verified that, as predicted analytically [1], only a few of the plane waves which compose the Gaussian spectrum may tunnel through such a bilayer. Consequently, a reflected wave remains present even after steady-state conditions are reached and a beam with a smaller angular spectrum tunnels through the bilayer even in the steady-state regime. This case has not been shown here for the sake of brevity.

However, in Fig. 7 we compare the time responses that are generated when a Gaussian beam with a waist $w_0 = 0.5\lambda_0$ impinges on a conjugate bilayer that is designed following Eq. (2) to have complete tunneling at the incidence angle $\theta_i = \pi/9 = 20°$, i.e., $\varepsilon_{ENG}(\omega_d) = -3\varepsilon_0$, $\mu_{ENG}(\omega_d) = 2\mu_0$, $\varepsilon_{MNG}(\omega_d) = 2\varepsilon_0$, $\mu_{MNG}(\omega_d) = -1.3\mu_0$, $d_{ENG} = d_{MNG} = \lambda_0/10$, with the one generated by the same beam impinging on the conjugate-matched bilayer shown in Fig. 6. From Fig. 7 one can see that steady-state is reached later in the conjugate-matched case than in the conjugate case, since more plane waves have to contribute to the resonance. Nonetheless, the tunneling is eventually complete (same amplitude and phase at the entrance and exit faces). In the conjugate case, on the other hand, steady-state is reached more quickly, but the field at the exit face is lower than at the entrance face. It is interesting to emphasize, moreover, how the fields at the interface get a higher value in the conjugate-matched case in comparison to the conjugate one. This is related again to the different numbers of plane waves that may actively contribute to the resonance. In other words, the conjugate matched resonance indeed shows a higher Q factor keeping fixed the thickness of the bilayer and the losses in the materials. (We remind the reader here that, as shown in [1], a given conjugate bilayer shows total tunneling at a single incident angle in the steady-state, represented by Eq. (2), but a sufficiently high transmission in a given angular region, whose broadness depends on the thickness of the bilayer and its other constitutive and geometrical parameters. This explains why the contribution from a superposition of plane waves to the resonance in the conjugate case is not represented by a single plane wave, but by a specific set of them).

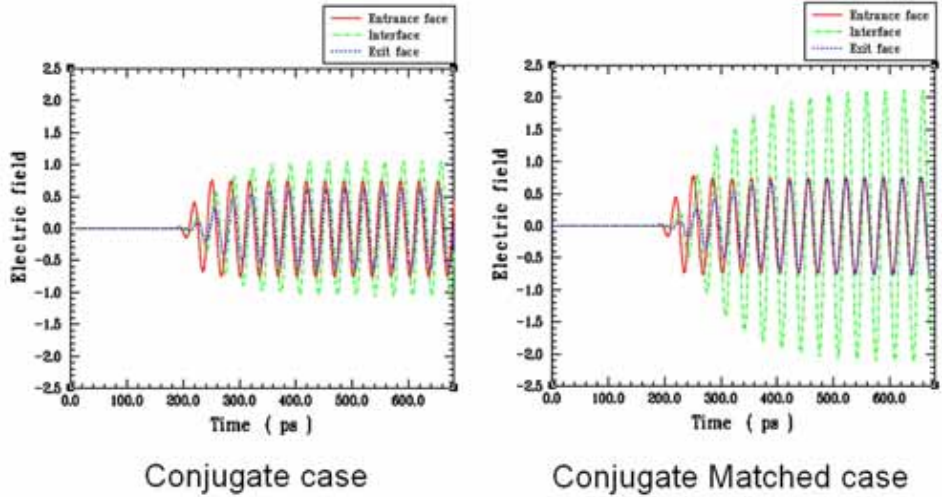

Fig. 7 – Time response for the same Gaussian beam (waist = 1.0) impinging on a conjugate bilayer and on a conjugate-matched one.

In Fig. 8 we have tested the possibility of employing the bilayer as a virtual image displacer, as proposed in [1]. We have considered a line source on one side of the bilayer, a distance $d = 20\Delta z = \lambda_0/5$ from it. It is clear from the Fig. 8 that after the time needed to reach steady-state, the phase and amplitude restoration at the exit side is complete, allowing for an observer placed on the exit side of the bilayer to "view" the line current as if it were closer than it actually is. Also the reflection on the source side is minimal once steady-state is reached.

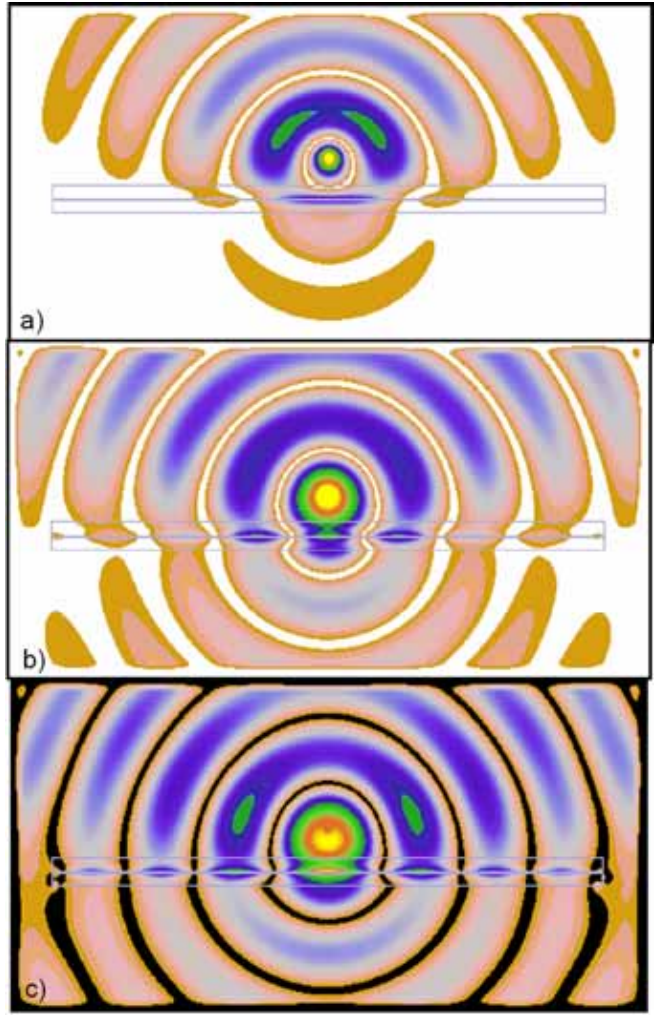

Fig. 8 – Three different snapshots in time (a) $t = 600\Delta t = 134.3\,ps$, (b) $t = 1000\Delta t = 223.9\,ns$, and (c) $t = 1900\Delta t = 425.4\,ps$, for the employment of the conjugate-matched pair of Fig. 2 as a near-field virtual image displacer.

As a final example, in Fig. 9 we have reported the steady-state regime electric field distribution for the same virtual displacer as in Fig. 8 when two sources with sub-wavelength spacing are placed close to the entrance side. The spacing between the sources is equal to $40\Delta z = \lambda_0/5$. You notice the expected large growth of the field at the ENG-MNG interface in the plot and how a blurry, but noticeable resolution of the sources at the back face is clearly visible. In the figure, the vertical lines correspond to the position of the two sources and the horizontal lines delimit the ENG and MNG layers.

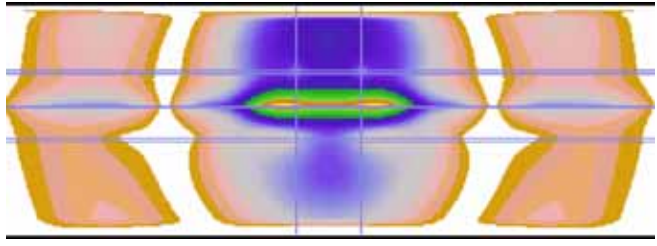

Fig. 9 – Snapshot in time for the electric field intensity at $t = 4850\Delta t = 1.715\,ns$ (in the steady state regime) for two electric current line sources separated by $40\Delta z = \lambda_0 / 5$ located $0.01\lambda_0$ away from the bilayer of Fig. 8.

## CONCLUSIONS

In this contribution, using the FDTD technique we have analyzed thoroughly in the time domain the anomalous resonant phenomenon arising when pairing together material slabs with opposite signs for the real parts of their constitutive parameters. Complete tunneling, total transparency, reconstruction of evanescent waves and sub-wavelength virtual imaging has been demonstrated numerically to occur after a reasonable time delay, even though each of the two slabs by itself would be essentially opaque to the incoming radiation. The effect works well even with transversally finite slabs and excitations, potentially leading to interesting applications for imaging tools. Physical insights and dependence of the time response to some of the parameters involved have been discussed.